%% file: main.tex
\documentclass{sig-alternate-per-Performance2018}

\usepackage{gianluca}

\usepackage{pdfpages}

\usepackage[colorinlistoftodos]{todonotes}

\usepackage{ulem}
\usepackage{endnotes}
\usepackage{natbib}
\setcitestyle{numbers,square}

\begin{document}
\conferenceinfo{Workshop on AI in Networks (WAIN) 2018}{~~~Toulouse, France}

{}

\title{A Deep Learning Strategy for Vehicular Floating Content Management}

\numberofauthors{3}
\author{
\alignauthor
Gaetano Manzo\\
Juan Sebastian Otalora \\
       \affaddr{University of Applied Sciences Western Switzerland (HES-SO)}\\
       \email{name.surname@hevs.ch}
\alignauthor
Marco Ajmone Marsan \\
       \affaddr{IMDEA Networks Institute, Spain}\\
       \affaddr{Politecnico di Torino, Italy}\\
       \email{ajmone@polito.it}
\alignauthor
Gianluca Rizzo \\
      \affaddr{University of Applied Sciences Western Switzerland (HES-SO)}\\
       \email{gianluca.rizzo@hevs.ch}
}


\maketitle
\input{abstract.tex}


\input{new_intro}

\input{system_model_new}

\input{prob_form.tex}
\input{algorithm}
\input{Num_eval}

\input{conclusions}

\bibliographystyle{ACM-Reference-Format}
\bibliography{main.bib}
\end{document}

%% file: abstract.tex
\begin{abstract}
Floating Content (FC) is a communication paradigm for the local dissemination of
contextualized information through D2D connectivity, in a way which minimizes the use of resources while achieving some specified performance target. Existing approaches to FC dimensioning are based on unrealistic system assumptions that make them, highly inaccurate and overly conservative when applied in realistic settings. 
In this paper, we present a first step towards the development of a cognitive approach to efficient dynamic management of FC. We propose a deep learning strategy for FC dimensioning, which exploits a Convolutional Neural Network (CNN) to efficiently modulate over time the resources employed by FC in a QoS-aware manner. Numerical evaluations show that our approach achieves a maximum rejection rate of $3\%$, and resource savings of $37.5\%$ with respect to the benchmark strategy.
\end{abstract}


%% file: new_intro.tex
\section{Introduction}
In the next generations of mobile internet connectivity, starting with 5G, new techniques to offload cellular networks will play a key role in coping with traffic peaks and to deliver quality of service and/or quality of experience (QoS/QoE) to end users.  Opportunistic communications are a prime candidate for cellular network offloading. Indeed, much of the traffic which is generated in such scenarios as Smart Cities, or autonomous coordinated driving, is local in scope and interest, and it is hence more efficiently delivered via direct exchanges between users. 

An interesting opportunistic communication scheme, for the local dissemination of information to end users through direct terminal-to-terminal connectivity, is Floating Content (FC)~\cite{floating}. 
The objective of FC is to spread content to the minimum amount of users within a defined area called Anchor Zone (AZ), which is sufficient for the content to persist over time in the AZ, while achieving a target performance. Typically, the performance parameter that characterizes the goodness of a FC scheme is the \textit{success ratio}, i.e. the average fraction of nodes with content \textit{entering} a given location (henceforth denoted as \textit{Zone of Interest} or ZOI). 

In FC, the purpose of constraining the opportunistic replication of a given content is to minimize the usage of resources (bandwidth, memory). Existing formulations (see \cite{floating} and references therein) perform a crude optimization, based on a coarse user partitioning on geographic criteria. Indeed, in almost all existing formulations~\cite{ours2017mobiworld,itc29} and~\cite{MONET2017}, the FC dimensioning problem boils down to finding the minimum AZ radius that guarantees a given target value of success ratio. 

The main limitation of these approaches is their being unfit for application in realistic settings. Indeed, the stationary assumption of the mobility patterns, and of uniformity of user distribution in space, enable the analytical evaluation of the fundamental performance trade-off in FC dimensioning. 
Such assumptions do not apply to most practical settings, in which mobility patterns exhibit features such as clustering and time correlation, and change continuously over time.
In particular, how to efficiently allocate resources (i.e., memory, bandwidth, and infrastructure support to content seeding) for an FC scheme when the content validity is short, or when the population of nodes in a given area and/or mobility patterns vary significantly over time, is still an open issue.\\
%
In this work, we present a first step towards addressing these issues. We propose a cognitive approach for dynamic management of FC schemes in vehicular scenarios. We consider settings with infrastructure support to FC, where the cellular network collects data on user mobility, and dynamically configures FC schemes. Our approach exploits a Convolutional Neural Network (CNN) to modulate over time the parameters of the FC scheme in order to achieve the target performance while minimizing a given cost function that accounts for the number of resources employed by the FC scheme. 
Through a numerical assessment we show that our approach achieves a maximum rejection rate of $3\%$ (i.e., the performance objective is not met only in $3\%$ of cases), resource savings of $37.5\%$ with respect to the benchmark strategy, and an accuracy of $89.7\%$ on the target message availability. 

The paper is structured as follows. Section~\ref{system} describes the system model, followed by the problem formulation in Section~\ref{prob}. Our deep learning algorithm is illustrated in Section~\ref{alg}. Finally, Section~\ref{concl} concludes the paper. 


%% file: system_model_new.tex
\section{System Model}
\label{system}
We consider a set of wireless nodes with transmission range $r$, moving on the plane according to a road grid. Nodes can model vehicles, or a combination of vehicles and pedestrians. We assume the road grid to be partitioned into a set $\mathcal{L}$ of \textit{road links}. 
The choice of such partition (and of the size of each road link) is related to a tradeoff between computational complexity and accuracy in the representation of the spatio-temporal mobility patterns.
We assume that two nodes come in contact when they are in the range of each other, i.e. when their distance is not larger than $r$. At any time, each node knows its exact position in space. 
%
\subsection{Cognitive FC operation}
In what follows, we describe the operation of the Cognitive Floating Content (CFC) communication paradigm.\\
We consider a finite time window, corresponding to the period of time during which content has to be made available over a portion of the considered road grid. The time window is partitioned into T intervals, each of duration $d_t$. We assume that at the beginning of the time window, the infrastructure has perfect knowledge of the main mobility parameters (e.g., mean node speed in each link, mean contact rate) for each interval in the time window. This is realistic as pedestrian and vehicular mobility patterns in urban scenarios exhibit periodic behaviors over time, and can thus be accurately predicted in advance. 

A CFC scheme is identified, for every road link $l$ and time interval $t=1,...,T$, by the \textit{content infectivity} $a_{l,t}$, and \textit{recovery rate} $b_{l,t}$ (with $a_{l,t}, b_{l,t}\in[0,1]$), as explained next. We assume such parameters are assigned by the system at the beginning of the time window. At the start of the first time interval, in every link of the road grid for which $a_{l,t}>0$, we assume that a given piece of content is present in at least one user in the link (this user is a \textit{seeder}). Such content is injected by the infrastructure (e.g., through cellular or WIFI communications), and it may be originated by the mobile nodes themselves (for instance, it can convey a warning about a road accident) or by an external source. Within each time interval $t$, whenever two nodes come in contact, if the content is present at only one of the two nodes, the content is transferred to the other node with probability $a_{l,t}$, where $l$ is the link where the sender node resides. Such transfer is subject to all limitations due, e.g. to capacity between the two nodes, to propagation effects such as fading, to interference.

When content is transferred, the receiver (residing on link $l'$) keeps it with probability $b_{l',t}$, and discards it otherwise. In general, whenever a node moves into the $l'$-th road link with content, it keeps it with probability $b_{l',t}$.

In this paper, we consider the ideal case in which nodes do not replicate content when it is not needed (i.e., when both nodes in contact already possess the content). We also assume that content exchanges are always unicast (one-to-one). Note however that it is possible to extend the CFC opportunistic communication paradigm to a multicast and multi-hop routing. Finally, since a study of an event-based mobility is out of the scope of this work, we assume that the mobility features are not affected by the diffusion of the floating content.

The resulting matrix $\mathbf{A}=\{a_{l,t},b_{l,t}\}$ completely describes a CFC scheme, by identifying the content replication and caching strategies over the whole time window. 

Following content seeding, if enough content replications have taken place, the content may persist over time in the road grid even when the seeder nodes have moved out of the road grid, or have discarded the content. We say in this case that the content \textit{floats}, i.e., it persists probabilistically in the considered scenario. This usually happens for a duration which is determined by the mobility pattern, and by the choice of the CFC parameters $\mathbf{A}$, among others. 

An important performance metric for CFC is the content \textit{availability}, i.e. the mean fraction of users with content in a time interval. 
As we previously mentioned, the goal of CFC is to make the floating content available in a subset $\mathcal{L}'$ of the links of the road grid, the \textit{Zone of Interest} or (ZOI), with a given mean availability. Specifically, such parameter, henceforth denoted as \textit{success ratio}, is the mean fraction of users in the ZOI with content. If $n_{l,t}$ ($n^c_{l,t}$) is the mean number of users (resp. mean number of users with content) on link $l$ during time interval $t$, the success ratio $\alpha_t$ in time interval $t$ is given by
\[
\alpha_t=\frac{\sum_{l\in\mathcal{L'}}n^c_{l,t}}{\sum_{l\in\mathcal{L'}}n_{l,t}}
\]
The target value $\alpha_0$ of the success ratio (typically constant in the whole time window), as well as the choice of size, shape, and location of the ZOI, depend entirely on application-level performance requirements.

In a CFC scheme, the infrastructure collects floating car (or user) data, and computes forecasts for the main mobility features in the whole time window. When required to establish an FC scheme, with a given ZOI and target success ratio, starting from its mobility forecasts it elaborates a CFC strategy, and it injects the coefficients $\mathbf{A}$ of such strategy in all the nodes of the grid.

%% file: prob_form.tex
\section{Formulation of the optimization problem}
\label{prob}
In this section, we formulate the problem of optimal CFC dimensioning. To this purpose, we define a cost function that is proportional to the number of resources employed by a CFC scheme. The first component of cost accounts for user memory on link $l$ at time $t$, and it is equal to the mean total number of users with content $n^c_{l,t}$, times the content size in bits, $D$.
The second component of cost accounts for the average amount of user bandwidth used to exchange content, and it is equal to the mean total number of users transmitting the content in a given time interval on a link. We denote it with $\gamma_{l,t}$.
Note that $\gamma_{l,t}$ depends, among other things, on content size, on available bandwidth, and on the distance between transmitter and receiver. Hence, the higher is the channel capacity, the lower is $\gamma_{l,t}$.\\
An optimal CFC scheme $\mathbf{A}$ is hence a solution of the following problem:
\begin{problem}[CFC resource optimization problem]\label{prob:1}
\begin{equation}\label{eq:objectivefunction}
\minimize_{\mathbf{A}}\;\;\sum_{l\in\mathcal{L}, t=1,...,T} \frac{(Dn^c_{l,t}+\beta\gamma_{l,t})d_t}{\sum_{t=1,...,T} d_t}
\end{equation}
	Subject to:
	\vspace{-0.15in}
	\begin{align}
	&\forall t,\gap\gap \alpha_t\geq \alpha_0\label{eq:availability_zoi}\\
	&\forall t, l\gap\gap 0\leq a_{l,t}\leq 1,\;\;\gap\gap 0\leq b_{l,t}\leq 1
	\end{align}
\end{problem}
Note that the cost function is the weighted sum of the cost components for each time interval, where each component's weight is the ratio between the interval duration and the duration of the whole time window. The coefficient $\beta\geq 0$ in the objective function is the factor that modulates the relative weight of the two cost components (the one on memory and the one on user bandwidth). 

A notable configuration of the CFC corresponds to the case in which $\forall \;l,t,\;a_{l,t}=1,\;b_{l,t}=0$. That is, to the case in which the content is replicated at every opportunity within the whole considered road grid, and never dropped. Such solution, which we denote as \textit{all-on}, is equivalent to making all resources in the system available to the CFC communication scheme, and typically it allows deriving the largest value of mean availability in the ZOI achievable with the given road grid. For these properties, the optimal cost of the solution(s) of Problem~\ref{prob:1} is always less than or equal to the cost of the all-on configuration.

In general, Problem $1$ cannot be solved efficiently, as $\forall\;l,t,\;n^c_{l,t}$ and $\gamma_{l,t}$ depend on $\mathbf{A}$ in ways which are hard to capture analytically without strong assumptions (e.g., stationary patterns mobility and uniformity of user distributions in space).

%% file: algorithm.tex
\section{A deep learning algorithm for efficient FC dimensioning}
\label{alg}
In this section, we describe our deep learning approach to solving Problem~\ref{prob:1}. Given a request of message spreading, and the forecast on traffic mobility, our algorithm selects the strategy $\mathbf{A}$ that achieves the performance targets required by the application (in terms of minimum success ratio) in a resource-efficient manner. 

The learning approach of choice has been the Convolutional Neural Network (CNN), a specialized type of neural network for processing data that has a known, grid-like topology \cite{goodfellow2016deep}. As we explained below, CNNs enable deep learning architectures to properly capture correlations between the spatial features of the road grid, the way in which the content spreads from one road link to another, and the way it persists in the road grid. 
%
\subsection{Training process}
The set of input parameters to the CNN algorithm is the \textit{link features vector} $\mathbf{P}=\{P_{l,t}\}=(\mathbf{P_m},\mathbf{P_c})$.
It consists in a set of parameters related to node mobility $\mathbf{P_m}$ and to content replication $\mathbf{P_c}$, associated to a given road link and time interval. In the present work, we have chosen as link features those parameters which are typically chosen to parametrize, in practical settings, the existing analytical models for FC performance \cite{ours2017mobiworld,itc29}. In particular for $\mathbf{P_m}$: the mean node speed, the mean number of nodes in a link. Whereas for $\mathbf{P_c}$ the mean rate of contacts in a link, and the mean duration of each contact. In addition, in $\mathbf{P_c}$ we have also considered those parameters that allow us to compute the cost function (\ref{eq:objectivefunction}), i.e. the mean number of nodes with content in a link, and and the mean number of concurrent content exchanges in a link. Through the communication feature $\mathbf{P_c}$ is possible to verify the conditions in \ref{eq:availability_zoi}. In case the target is not achieved, the strategy $\mathbf{A}$, for the respective $\mathbf{P_c}$ and $\mathbf{P_m}$, is set to \textit{off} (i.e., it is not possible to store or spread the message in the whole roadmap). This data pre-processing helps the machine to understand which configurations do not satisfy the condition in \ref{eq:availability_zoi}. Finally, the input $\mathbf{P}$ and the respective label $\mathbf{A}$, are ready for the learning process. 

The learning process uses a CNN to relate the link features $\mathbf{P}$ to their labels $\mathbf{A}$. The process used to build a training set for the CNN, i.e. to train the CNN to distinguish those inputs which lead to a strategy $\mathbf{A}$, which satisfies the constraint on success ratio \eref{eq:availability_zoi}), is as follows. As we have already stated, we assume the infrastructure constantly collects floating car data over the whole road grid, and data about communications. Specifically, in addition to computing the mobility parameters $\mathbf{P_m}$, in the link features vector, the infrastructure records each event of nodes coming in contact (i.e., within transmission range of each other) and the duration of each contact.

Given such historical information of contact patterns, and a strategy $\mathbf{A}$, numerical simulation allows computing the link features $\mathbf{P_c}$ associated with $\mathbf{P_m}$ and $\mathbf{A}$, as well as deciding whether $\mathbf{A}$ is a feasible strategy, i.e. whether $(\mathbf{P},\mathbf{A})$ satisfies the condition \eref{eq:availability_zoi}. The final training set consists of $N$ couples of matrices $(\mathbf{P},\mathbf{A})$, where the size $N$ is chosen as a compromise between computational load (and hence time to complete the CNN training) and accuracy of the output. 

As a result, thanks to the periodicity of the mobility patterns in a city, when a request is made for a content to be diffused by CFC in a given time window, for a given ZOI and with a given target success ratio, the infrastructure builds a training set, trains the CNN, inputs to the trained CNN the parameters $\mathbf{P_m}$ forecasted for the whole observation window, and obtains in this way the content diffusion strategy $\mathbf{A}$. Since multiple $\mathbf{A}$ are associated to the same $\mathbf{P}$, the CNN output $\mathbf{A}$ is the one that minimize \ref{eq:objectivefunction}. Note that, as mentioned above, the output $\mathbf{A}$ can be \textit{all-off} if the conditions in \eref{eq:availability_zoi} are not satisfied, i.e. it is not possible to spread or store in the whole roadmap.


\subsection{Model Architecture}
The Convolutional Neural Network (CNN) architecture enables capturing both intra-link and inter-link relationships between features. Indeed, in the CFC dimensioning problem, the information about proximity or relative position between links is not part of the link features, due to the complexity which including such information would imply on the algorithm. Moreover, the correlations between features of different links are not due only to proximity. Rather, they are also the result of wireless propagation effects and, most importantly, of spatio-temporal patterns in node mobility (e.g. typical patterns of vehicular traffic, in terms of sequence of link traversals). Unlike "non-deep" machine learning approaches, the significant features extracted by the CNN are not only related to a single link but they are associated with those links within the kernel. As a result, the characteristics extracted from each layer of the CNN do not consist only in local correlations. It also includes the long-spatial relations between links, due to the specific spatio-temporal patterns of the mobile nodes that typically have a strong impact on CFC performance.

\begin{figure}[t]
 \begin{center}
    \includegraphics[width=\linewidth]{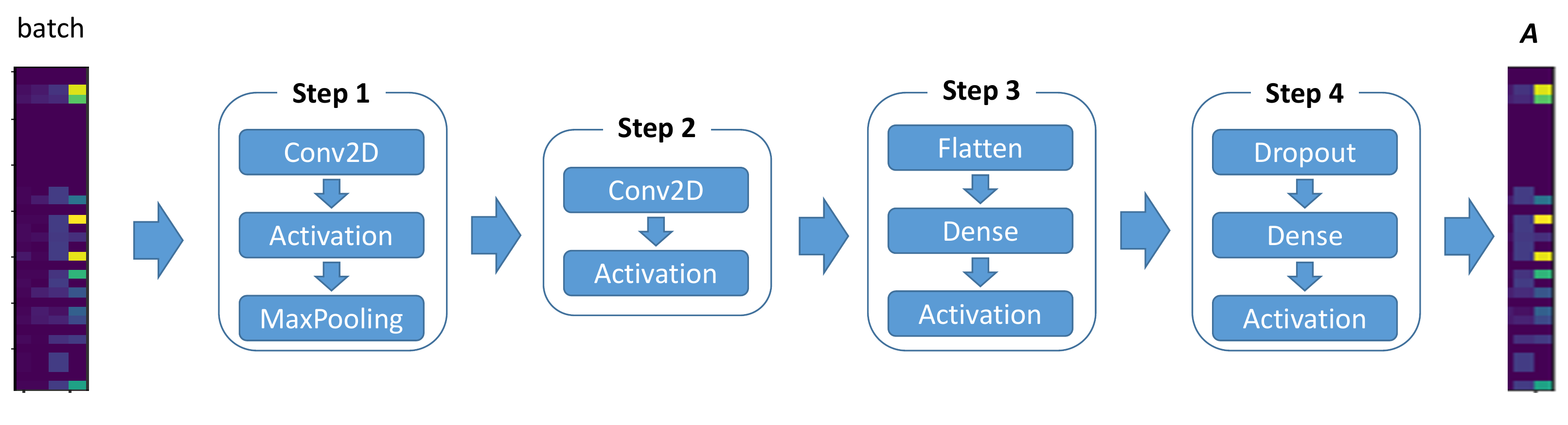}
\caption{\small Structure of the Convolutional Neural Network.}
\label{fig:cnn_arc}
\end{center}
\normalsize
\vspace{-0.2in}
\end{figure}

The detailed structure of the CNN architecture adopted is given in Figure~\ref{fig:cnn_arc}. The spatio-temporal correlations between links, are captured by the non-linear characteristics learned in these layers. From one layer to the next, these characteristics are then aggregated by applying the convolution, non-linearity and max-pooling operations. 
 

%% file: Num_eval.tex
\section{Numerical Evaluation}
\label{N-e}
In this section, we evaluate the performance of our proposed approach in an artificial road grid setting. In this scenario, our approach has been trained, validated and assessed.
We have considered a Manhattan road grid composed of square blocks of side $150$m, for a total of $35$ road links. Nodes enter the grid from links at the border, with an arrival rate equal to $3$ nodes/s, equal for all links. 
The ZOI has been chosen at the center of the grid. The target success ratio $\alpha_0$ has been set to $0.9$.
We considered $\beta$$=$$1$, a channel bandwidth of 4 MHz, a content size equal to $4$ MB, and the channel capacity resulting from Shannon's formula. 

The training set has been built as follows. The road link features have been measured over a time interval of $3750$ s, using a sampling time of $1$ s. We considered 
$4\times10^6$ couples $(\mathbf{P},\mathbf{A})$. 
A ten-fold cross-validation has been performed to avoid overfitting and reinforce model stability over unknown inputs. A set of unknown couples  $(\mathbf{P},\mathbf{A})$ has been used for testing. Specifically, we have used the mobility link features as input, the communication link features for application validation, and $\mathbf{A}$ as ground truth. Each time interval has a duration of one hour, as a reasonable trade-off between computational load and application performance.
We implemented the proposed architecture in the Keras deep learning framework and trained it using a Titan Xp GPU. 
%

\begin{figure}[t]
 \begin{center}
    \includegraphics[width=0.85\linewidth]{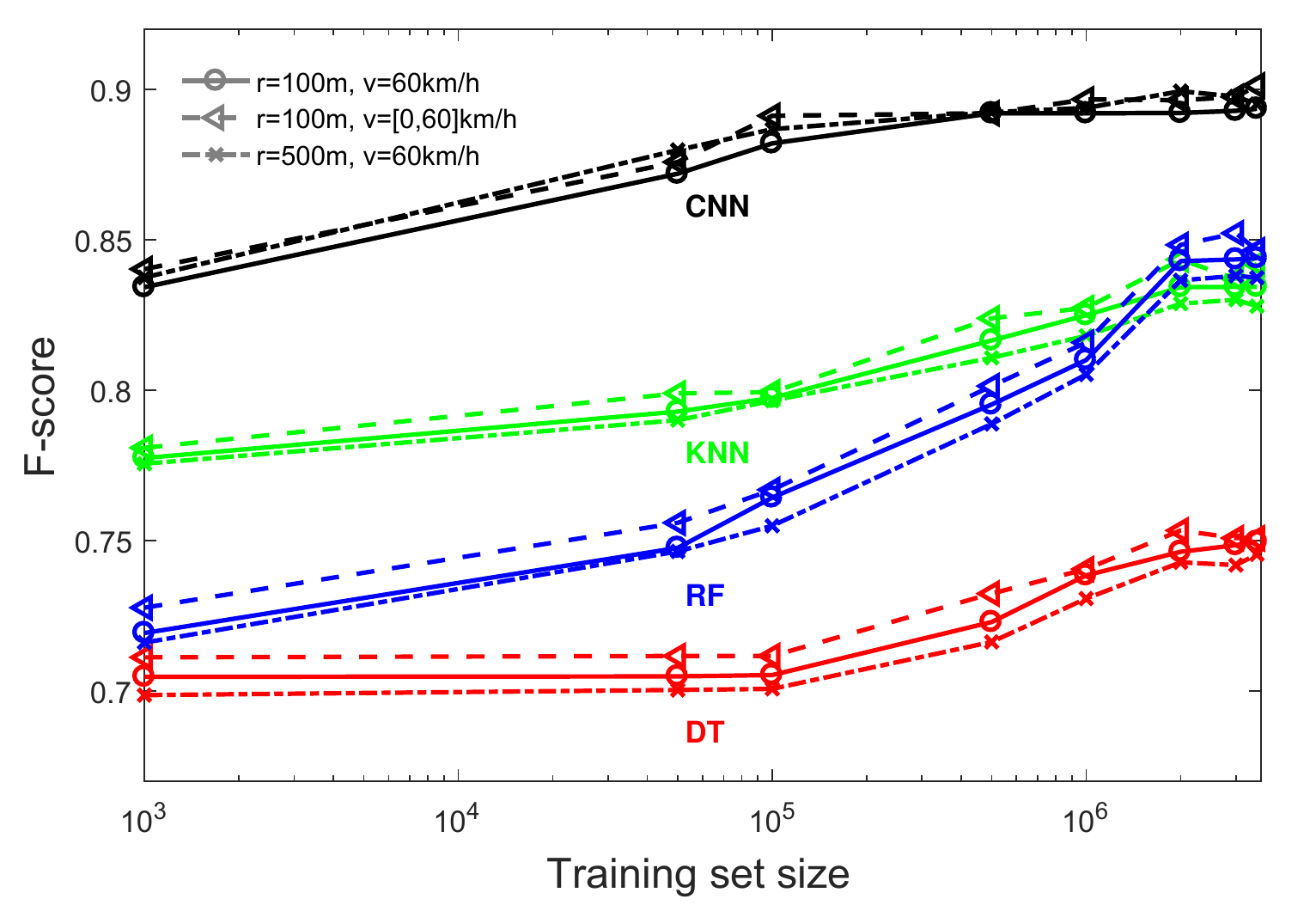}
\caption{\small F-score versus size of the training sample, for our CNN algorithm, as well as for K-Nearest Neighbor (KNN), Decision Tree (DT), and Random Forest (RF), for the three scenarios considered. All curves are with a $98\%$ confidence interval of $7\%$}
\label{fig:fscore}
\end{center}
\normalsize
\vspace{-0.3in}
\end{figure}

For evaluation of the accuracy of our algorithm, in Figure~\ref{fig:fscore} we have plotted the F-score versus the training set size, in three different combinations of transmission radius and of speed (constant, and equal to $60$ km/h, or uniformly distributed in the interval $[0, 60]$ km/h).
In each of these scenarios, we have also considered the performance of the other main algorithms for multi-label classification problems, i.e., K-Nearest Neighbor (KNN), Decision Tree (DT), and Random Forest (RF). Note that, for each algorithm, all scenarios are tested using over $1.5\times 10^5$ registers of the test set. In order to measure model accuracy, we limited the number of possible strategies. Each road link is classified in the range of $(0,10)$ for storing and replication strategy level.  

These results show that for all sizes of training set and in all considered scenarios, our deep learning approach substantially outperforms the other algorithms in terms of accuracy. In particular, our approach achieves a very high accuracy even with small training sets. This confirms that for the problem of efficient dimensioning of a CFC scheme, the ability of CNN to capture the impact that the features of a link have on the content replication performance in other links, is the key feature for achieving a satisfactory performance. Another indication of this is given by the fact that, by increasing the transmission radius from $100$ m to $500$ m, the relative accuracy of the CNN approach improves with respect to the other approaches. Indeed, by increasing the transmission radius, we increase the amount of content exchanges between links, which are close in space, and hence the impact, which such coupling has, on the overall performance of a CFC scheme. 
Note that, the scenario with variable speed, by adding one more feature to the learning process, brings to a higher \textit{F-score} in all algorithms. 

A key aspect, of any learning approach to FC dimensioning, is that even approaches which exhibit high accuracy may produce configurations which are not feasible, i.e. which do not satisfy the target performance in terms of minimum success ratio. Table~\ref{tab:output} shows, for all considered scenarios and algorithms, the probability that the output of the learning processes does not satisfy constraint \eref{eq:availability_zoi}. For our approach, the rejection probability is around $3\%$, and an order of magnitude lower than those of other approaches. Indeed, CNN tends to decrease the \textit{false positive} predictions rather than the \textit{false negative} ones. This is due to the fact that, for any input, our CNN always starts from the all-on configuration as initial state.
\begin{table}[htbp]
\footnotesize
\begin{center}
\caption{Rejection probability for the algorithms considered, with a $98\%$ confidence interval of $6\%$. Training set size: $1.5\times10^5$.}
\begin{tabular}{ c|c c c}

Tx radius [m]  & $100$ & $100$ & $500$ \\
Speed [Km/h] & $60$ & $[0,60]$ & $60$\\
  \hline
  
  CNN &\textbf{0.029} & \textbf{0.031} & \textbf{0.032}\\
  
  \hline
  KNN &0.274 & 0.287& 0.284\\
  \hline
  RF &0.324 & 0.320& 0.333\\
  \hline
  DT &0.347 & 0.345& 0.349
  \label{tab:output}
\end{tabular}
\end{center}
\normalsize
\vspace{-6mm}
\end{table}
%

A crucial performance parameter of our CNN approach is the percentage of resource cost saved with respect to the all-on configuration, in the scenario with $r=100m$, speed $60km/h$. The all-on configuration is considered as a reference, as it is the only configuration that by hypothesis is always feasible (if this is not the case, then there is no feasible solution for Problem~\ref{prob:1} for the given choice of road grid), and it represents therefore the fallback configuration to be adopted when other approaches fail.
In such conditions, our approach achieves $37.5\%$ of resource savings, hence substantially improving over trivial dimensioning methods. Note that, the percentage of resource savings increases drastically, if the mobility is not uniform unlike the above case.
A crucial aspect of our learning approach is the computational load required for building the training set and for training the CNN, as these computations are performed after the request for setting up a CFC scheme is formulated. Indeed, the training set cannot be precomputed, due to the many variables from which it depends (configuration and position of the ZOI, target success ratio, start time and duration of the time window). In addition, while for some applications a CFC can be planned, leaving enough time for training (e.g., when the content to float is an advertisement about a sale), for such applications as car accident notification, or medical emergency, a quick and effective deployment of the CFC may be necessary for acceptable performance.
In our evaluations, we also evaluate the training and testing times without a GPU. Using a I7 desktop PC with 16 GB of RAM, the two computational steps (training set buildup and CNN training) together took $3$ s for a training set of $1000$ elements, and about $7$ min for $10^6$ elements, while the computation of the CFC strategy $\mathbf{A}$ always took less than $1$ s at inference time.

%% file: conclusions.tex
\section{Conclusions}
\label{concl}
In this work, we have outlined a deep learning strategy for efficient FC dimensioning. It exploits a Convolutional Neural Network (CNN) to modulate over time the parameters of the FC scheme. In the continuation of this work, we plan to assess our approach in realistic settings with measurement-based traces. Moreover, we will extend it to the case in which mobility patterns are influenced by the spreading of the floating content (e.g., traffic jam notification).

